\documentclass{article}
\usepackage{spconf,amsmath,graphicx,mathtools,nth,hhline,multirow,color,cite}
\usepackage{url}

\newcommand{\vF}{\mathbf{F}}
\newcommand{\vW}{\mathbf{W}}

\newcommand{\vZ}{\mathbf{Z}}
\newcommand{\vX}{\mathbf{X}}

\title{A Streamlined Encoder/Decoder Architecture for Melody Extraction}
\name{Tsung-Han Hsieh$^{1,2}$, Li Su$^
3$ and Yi-Hsuan Yang$^2$}
\address{$^{1}$ Data Science Degree Program, National Taiwan University, Taiwan \\
$^{2}$ Research Center for IT Innovation, Academia Sinica, Taiwan \\
$^{3}$ Institute of Information Science, Academia Sinica, Taiwan \\
\tt \{bill317996,yang\}@citi.sinica.edu.tw, lisu@iis.sinica.edu.tw}

\begin{document}

\maketitle
\begin{abstract}
Melody extraction in polyphonic musical audio is  important for music signal processing. In this paper, we propose a novel streamlined encoder/decoder network that is designed for the task. 
We make two technical contributions. First, drawing inspiration from a state-of-the-art model for semantic pixel-wise segmentation, 
we pass through the pooling indices between pooling and un-pooling layers to localize the melody in frequency.
We can achieve result close to the state-of-the-art with much fewer convolutional layers and simpler convolution modules. Second, we propose a way to use the bottleneck layer of the network to estimate the existence of a melody line for each time frame, and make it possible to use a simple argmax function instead of ad-hoc thresholding to get the final estimation of the melody line. Our experiments on both vocal melody extraction and general melody extraction validate the effectiveness of the proposed model.
\end{abstract}
\section{Introduction}
\label{sec:intro}

Melody extraction is the task that aims to estimate the fundamental frequency (F0) of the dominant melody.
Automatic melody extraction has been an active topic of research in the literature, since it has many important downstream applications in music analysis and retrieval \cite{salamon14spm,kroher16taslp,bittner2017pitch,beveridge18pm}.

Lately, many deep neural network architectures have been proposed for melody extraction  \cite{rigaud2016singing,kum2016melody,bittner17ismir,su2018vocal,lu18ismir}. 
The basic idea of such neural network based methods is to use the neural nets to learn the mapping between a matrix that represents the input audio and another matrix that represents the melody line. For the input, it is usually a time-frequency representation such as the spectrogram, 
which can be viewed as an $F \times T$ real-valued matrix, where $F$ and $T$ denote the number of frequency bins and time frames, respectively. For the output, it is another $F \times T$ matrix but this time it is a binary matrix indicating the F0 of the melody line for each frame. We only consider music with a single melody line in the music, so at most one frequency bin would be active per frame. It is also possible that there is no melody for some frames. From the training data, we have a number of such input and output pairs. We can use the difference between the target output and the predicted one to train the neural net in a supervised way.

Existing work has shown that using the neural nets to learn the nonlinear mapping between audio and melody leads to promising result. However, there are two issues that require further research. First, as it is easier for a neural net to deal with continuous values, the output of most existing models (if not all) is actually an $F \times T$ real-valued matrix, not a binary one. This is fine for the training stage, since we can still use cost functions such as cross entropy to measure the difference between a real-valued matrix (the estimated one) and a binary matrix (the groundtruth). However, for the testing stage, we still need to \emph{binarize} the output of the neural net. This binarization cannot be easily achieved simply by picking the frequency bin with the maximal activation per frame, because this would lead to false positives for frames that do not have melody. Therefore, most existing methods have to use a threshold whose value is empirically determined in a rather ad-hoc way for binarization.
\cite{bittner17ismir,lu18ismir}

The second issue is that existing models that lead to state-of-the-art result in melody extraction benchmark datasets may be overly complicated.
For example, the model presented by Lu and Su \cite{lu18ismir} uses in total 45 convolution or up-convolution layers, using residual blocks for the convolution modules and a sophisticated spatial pyramid pooling layer. 
The goal of this paper is to propose a streamlined network architecture that has much simpler structure, and that does not need additional post-processing to binarize the model output. With a simple structure, we can better interpret the function of each layer of the network in generating the final result. We hope that the network can have accuracy that is comparable with, if not superior to, the state-of-the-art models.

We make two technical contributions to realize this. First, following Lu and Su  \cite{lu18ismir}, we use an encoder/decoder architecture
to learn the audio-to-melody mapping. But, while they use the skip connections to pass the output of the convolution layers of the encoder to the up-convolution layers of the decoder, we propose to add links between the pooling layers of the encoder and the un-pooling layers of the decoder, and pass along the ``pooling indices''\cite{segnet}. While the skip connections they use will be short paths for gradient propagation, there is no trainable weights in pooling and un-pooling layers. We argue from a functional point of view that our method makes it easier for the model to localize the melody.
Second, we propose to use the bottleneck layer of the network to estimate the existence of melody per frame, and design a way such that we can simply use argmax to binarize the output.
The final model has in total only 7 convolution or up-convolution layers.

\begin{figure}
\centering
\includegraphics[width=\columnwidth]{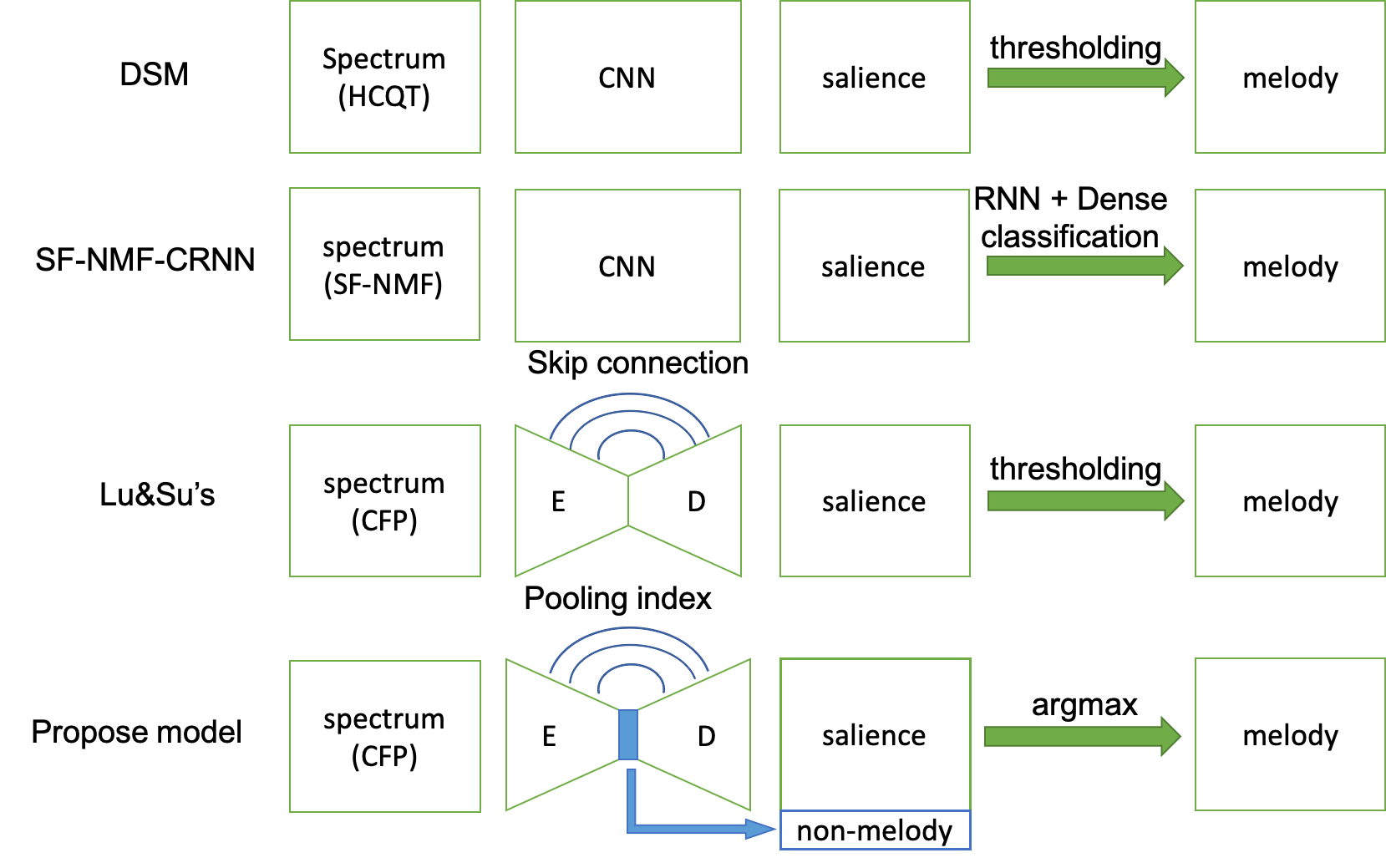}
\caption{Comparison of the network architecture of SF-NMF-CRNN \cite{basaran2018CRNN}, DSM \cite{BittnerDeepSalience17}, Lu \& Su's model \cite{lu18ismir}, and the proposed model. (Notation---`E': encoder, `D': decoder.)}
\label{fig:comparison}
\end{figure}

\section{Related work}
\label{sec:related}

We show in Fig. \ref{fig:comparison} the network architectures of three previous methods that are proposed lately. 
The first one is the deep salience model (DSM) proposed by Bittner \emph{et al.} \cite{bittner17ismir}. It uses a convolutional neural network (CNN) that takes a time-frequency representation of music as the input, and generates a salience map as output for estimating the melody. 
Finally, they apply a threshold to the salience map to get the binary melody estimate.
The second one is the SF-NMF-CRNN model proposed by Basaran \emph{et al.} \cite{basaran2018CRNN}.
Instead of thresholding, it learns recurrent  and dense layer to binarize the frequency map.
Another model presented by Lu and Su \cite{lu18ismir}, which is based on the DeepLabV3+ model \cite{Chen2018DeepLabV3+}, shows that better result for vocal melody extraction can be obtained by an encoder/decoder architecture with skip connections.
This model also uses thresholding to binarize the result. 

The thresholding operation
can be found in many music-related tasks.
It can be done with a fixed threshold, an adaptive threshold \cite{lu18ismir}, or
other advanced methods  \cite{dong18ismir,southall18ismir}.

\begin{figure}
\centering
\includegraphics[width=\columnwidth]{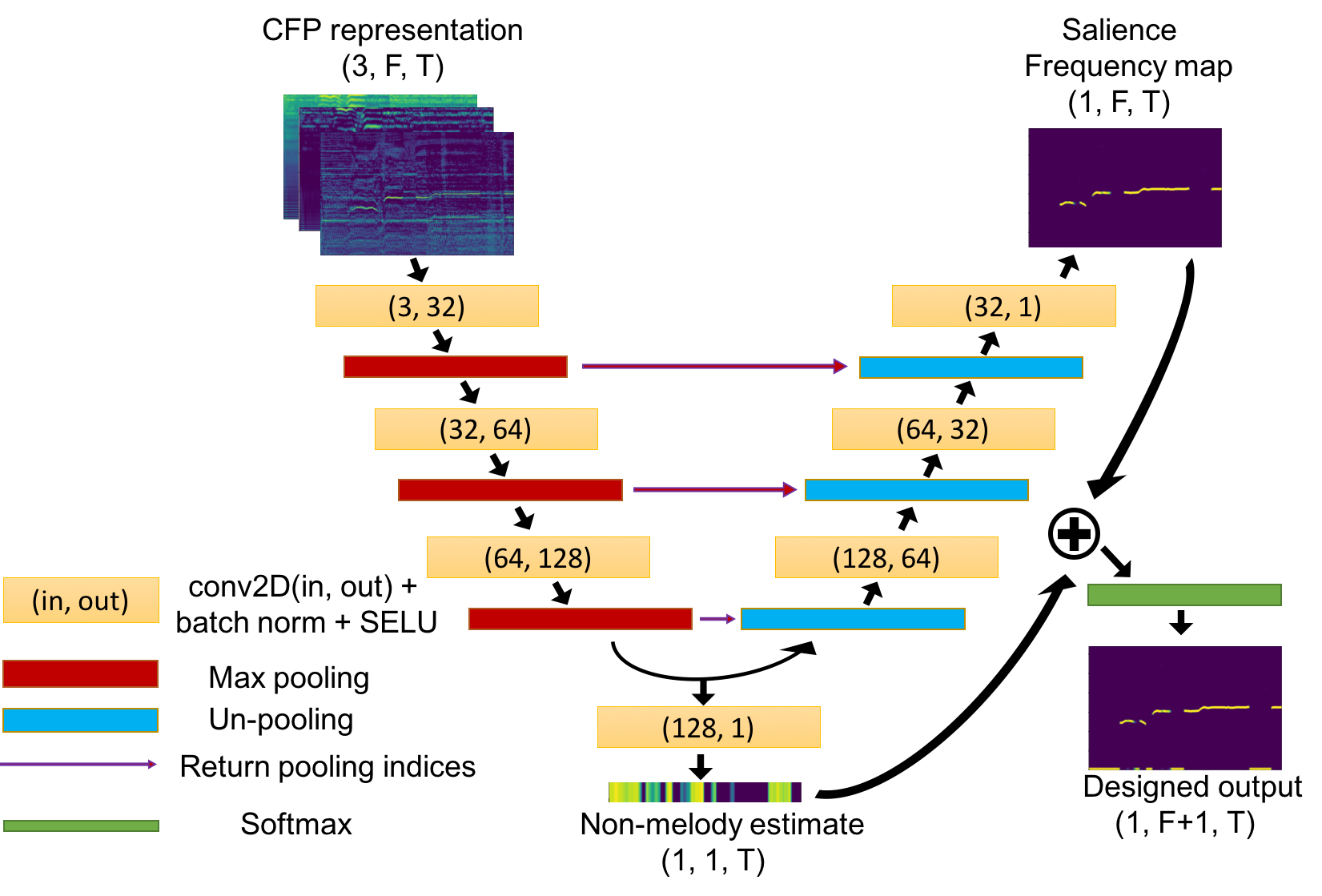}
\caption{Details of the proposed model. We
detect non-melody activity as a sub-target at the bottleneck layer and concatenate it with the output of the decoder, the salience frequency map.}  
\label{fig:model}
\end{figure}

\section{Proposed Model}
\label{sec:Method}
The system overview is given in Fig. \ref{fig:model}. It has a simple  encoder/decoder architecture. For
the encoder, we use three convolution layers and three max pooling layers. The output of the encoder is taken as the input by two separate branches of layers. The first branch is simply the decoder that uses three up-convolution layers and three un-pooling layers to estimate the 
salience frequency map. 
The second branch uses one convolution layer to estimate
the existence of melody per frame, leading to the ``non-melody'' estimate in the bottom.
Finally, the salience map and the non-melody estimate are then concatenated (along the frequency axis), 
after which we get a binary-valued estimate of the  melody line with a simple softmax layer.
We give more details of the network below.

\subsection{Model Input}
While the model can take any audio representation as the input, we choose to use  the Combined Frequency and Periodicity (CFP) representation \cite{su2015combining}.
It contains three parts:  the power-scaled spectrogram, generalized cepstrum (GC) \cite{kobayashi1984spectral,tokuda1994mel}
and generalized cepstrum of spectrum (GCoS) \cite{su2017HSP_DNN}.
The latter two are \emph{periodicity} representations that have been shown useful to multi-pitch estimation (MPE) \cite{peeters2006music}. 
Given $\vX$, the magnitude of the short-time Fourier transform (STFT) of an input signal, GC and GCoS can be computed as:
\begin{align}
\vZ_{\text{S}}[k,n]&:=\sigma_{0}\left(\vW_f\vX\right)\,, \label{eq: specs}\\ \vZ_{\text{GC}}[q,n]&:=\sigma_{1}\left(\vW_t\vF^{-1}\vZ_{\text{S}}\right)\,, \label{eq: ceps} \\ 
\vZ_{\text{GCoS}}[k,n]&:=\sigma_{2}\left(\vW_f\vF\vZ_{\text{GC}}\right)\,, \label{eq: gcos}
\end{align}
where $\vW_f$ and $\vW_t$ are high-pass filters for removing the DC terms, $\vF$ an DFT matrix  and $\sigma_i$ activation functions \cite{su2015combining}.
\begin{figure}
\centering
\includegraphics[width=\columnwidth]{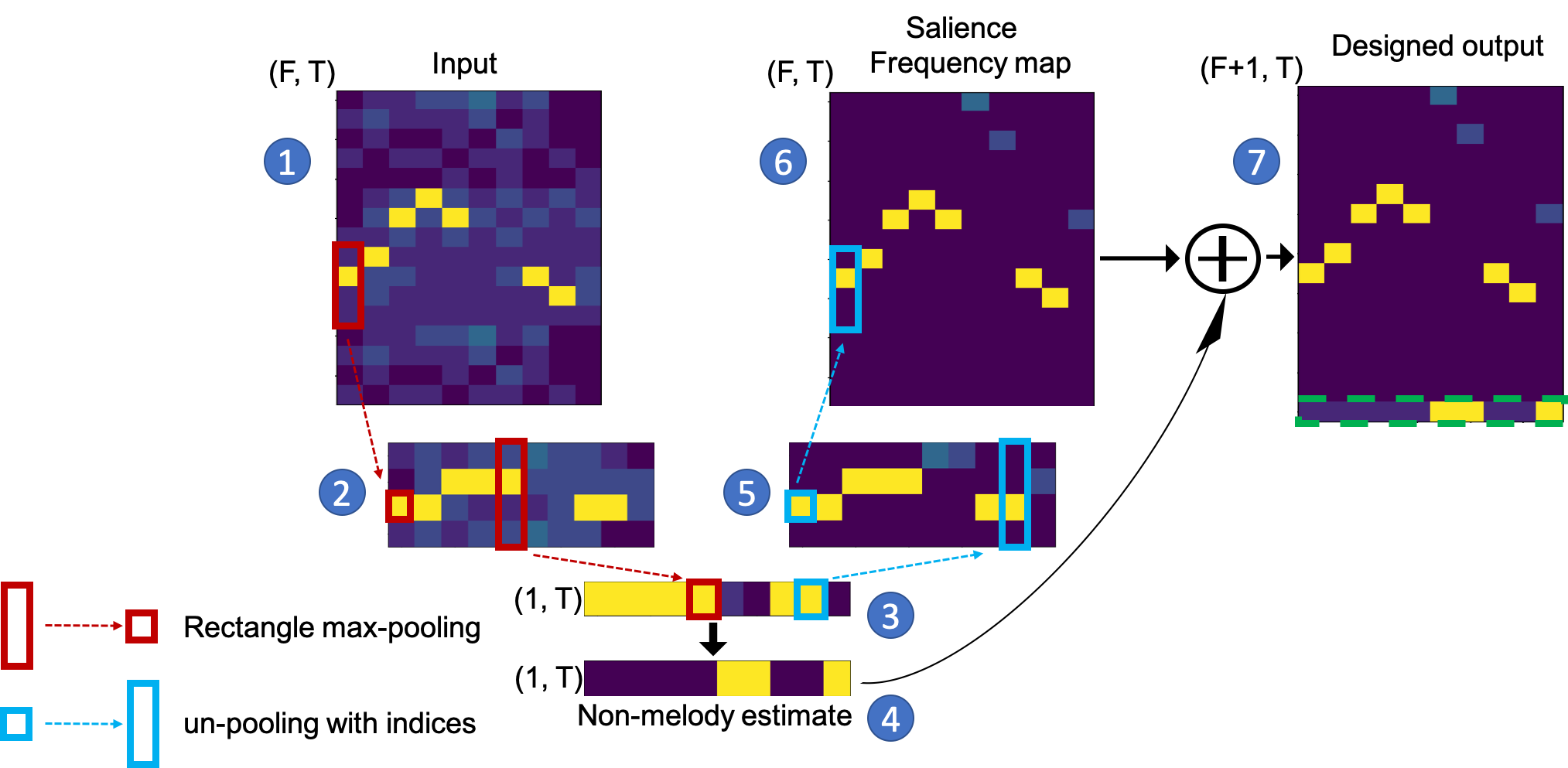}
\caption{An illustration of our model in action. 
For time frames with  melody notes, their values would be close to 1 (bright yellow) in the output of the last encoding layer (\textcircled{\raisebox{-0.9pt}{3}}), but close to 0 (dark blue) in the output of the non-melody detector (\textcircled{\raisebox{-0.9pt}{4}}). We ``reverse the bits" here because we can then concatenate \textcircled{\raisebox{-0.9pt}{4}} with \textcircled{\raisebox{-0.9pt}{6}} so that a simple argmax on \textcircled{\raisebox{-0.9pt}{7}} can tell us whether there is a melody note and where it is. }


\label{fig:demo}
\end{figure}

\subsection{Encoder and Decoder}
The design of the encoder and decoder represents the first technical contribution of this work.
As depicted in Fig. \ref{fig:model}, we use simple convolution/up-convolution and pooling/un-pooling layers in our model. Moreover, we 
pass the pooling indices between the pooling and un-pooling layers.  

The design is motivated by SegNet \cite{segnet}, a state-of-the-art model for semantic pixel-wise segmentation of images. We found that melody extraction is  similar to image segmentation in that both tasks require learning the mapping between a real-valued, dense matrix 
and a binary-valued, relatively sparser matrix. 
For melody extraction, the target output is indeed sparse---we only have at most one active entry per column (i.e. per time frame). Therefore, we like to test the idea of SegNet and use
pooling indices to inform the un-pooling layers  the exact entries picked by the pooling layers in the encoding process. 
This makes it easier for the decoder to localize the melody in frequency.
This is illustrated in Fig. \ref{fig:demo}.

In each convolution block, we use only one convolution layer with batch normalization and scaled exponential linear units (SELU) \cite{Klambauer17arxiv} as the activation function. The convolution kernel size is (5,5) with padding size (2,2) and stride size (1,1). 
For the max-pooling layer, we use kernel size (4,1) and pool only along the frequency dimension. The feature map at the bottleneck of the network is a $128 \times T$ matrix. 


\subsection{Non-melody Detector and  ArgMax Layer}
The design of the non-melody detector represents the second technical contribution of this work. 
As depicted in Fig. \ref{fig:model}, we learn one additional convolution layer that converts the $128 \times T$ matrix into a $1 \times T$ vector. 
This vector is then concatenated with the salience map to make an $(F+1) \times T$ matrix, where the last row corresponds to this vector (see Fig. \ref{fig:demo} for an illustration).
We then use the argmax function to pick the entries with the maximal value per time frame and return the melody line with the following rule---\emph{if the argmax is the $F+1$ entry for a frame, we consider that there is no melody for that frame}. 
In this way, the output of the model is an $F \times T$ binary matrix with only one or no active entry per frame.

In the model training process, the model output would be compared with the groundtruth output to calculate the loss and to update the network parameters. Therefore, according to our design, the  convolution layer we just mentioned would be ``forced'' to learn whether there is a melody for each frame. Moreover, the frames without melody would tend to have high activation (close to `1'), whereas those with melody would have low activation (close to `0'), as shown in Fig. \ref{fig:demo}. This is why we call this branch the non-melody detector.

We can view the non-melody detector as a singing voice detector \cite{lehner14icassp, schlueter15ismir,Danieljointdetect} when the task is to detect the vocal melody. 
But, our design is for general melody extraction, not only for vocal melody extraction.

The argmax layer is significant in that we do not need a separate, postprocessing step to discern melody/non-melody frames and to binarize the model output. The non-melody detection and binarization are built-in and trained together with the rest of the network to optimize the accuracy of melody extraction. 
To our best knowledge (also see Section \ref{sec:related}), there is no such a model in the literature.

The argmax layer is not a general solution for any music-related tasks that require binarization. For example, in MPE \cite{su2015combining,peeters2006music} there are usually multiple active entries per frame.



\subsection{Model Update}

While Lu and Su \cite{lu18ismir} use the 
focal loss \cite{Lin2017focal_loss} to deal with the sparsity of melody entries, we find our model works well with a simple loss function---the binary cross entropy  between the estimated melody and the groundtruth one. Model update is done with mini-batch stochastic gradient descent (SGD) and the Adam optimizer. 
The model is implemented using PyTorch. 
For reproducibility, we share the source code at
\url{https://github.com/bill317996/Melody-extraction-with-melodic-segnet}.




\section{Experiment}
\label{sec:exp_setup}

\subsection{Experimental Setup}

We evaluate the proposed method on general melody extraction for one dataset, and on vocal melody extraction for three datasets.
For \textbf{general melody extraction}, we use the MedleyDB dataset \cite{bittner14ismir}. Specifically, we use the ``melody2'' annotation, which is the F0 contours of the melody line drawn from multiple sound sources. Following \cite{basaran2018CRNN}, among the 108 annotated songs in the dataset, we use 67 songs for training, 14 songs for validation and 27 songs for testing.

For \textbf{vocal melody extraction}, we use the MIR-1K dataset
\footnote{https://sites.google.com/site/unvoicedsoundseparation/mir-1k}
and a subset of MedleyDB for training. The former contains 1,000 Chinese karaoke clips, whereas the latter contains 48 songs where the vocal track represents the melody. 
The testing data are from three datasets: 12 clips from ADC2004, 9 clips from MIREX05,\footnote{https://labrosa.ee.columbia.edu/projects/melody/} and 12 songs 
from MedleyDB. 
We set the training and testing splits of MedleyDB according to \cite{lu18ismir}.
There is no overlap between the two splits.


We compare the performance of our model with the three state-of-the-art deep learning based methods \cite{BittnerDeepSalience17,basaran2018CRNN,lu18ismir} described in Section \ref{sec:related}. Moreover, to validate the effectiveness of the non-melody detector branch, we implement an \emph{ablated} version of our model that removes the non-melody detector. For binarization of this method, we run a grid search to find the optimal threshold value using the validation set.

Following the convention in the literature, we use the following metrics for performance evaluation: overall accuracy
(OA), raw pitch accuracy (RPA), raw chroma accuracy
(RCA), voicing recall (VR) and voicing false alarm (VFA).
These metrics are computed by the  \texttt{mir\_eval} \cite{mireval} library with the default setting---e.g., a pitch estimate is considered correct if it is within 50 cents of the groundtruth one.
Among the metrics, OA is often considered more important.

To adapt to different pitch ranges required in vocal and general melody extraction, we use different hyperparameters in computing the CFP for our model. For vocal melody extraction, the number of frequency bins is set to 320, with 60 bins per octave, and the frequency range is from 31 Hz (\texttt{B0}) to 1250 Hz (\texttt{D\#6}). For general melody extraction, the number of frequency bins is set to 400, with 60 bins per octave, and the frequency range is from 20 Hz (\texttt{E0}) to 2048 Hz (\texttt{C7}). 
Moreover, since we use more frequency bins for general melody extraction, we increase the filter size of the third pooling layer of the encoder from  (4,1) to (5,1) for this task. 

We use 44,100 Hz sampling rate, 
2,048-sample window size, and 256-sample hop size for computing the STFT.
Moreover, to facilitate training the model with mini-batches, we divide the training clips into fixed-length segments of $T=256$ frames, which is nearly 1.5 seconds.
According to our implementation, the model training can converge within 20 minutes with a single GTX1080ti GPU.

\vspace{-2mm}
\subsection{Result}
Table \ref{tab:mdb main} first lists the performance of vocal melody extraction for three datasets. 
We see that the proposed model compares favorably with DSM \cite{BittnerDeepSalience17} and Lu \& Su's model \cite{lu18ismir}, leading to the highest OA for the ADC 2004 and MedleyDB datasets. 
In particular, the proposed model outperforms the two prior arts greatly for 
MedleyDB, the most challenging dataset among the three. 
We also see that the proposed method outperforms DSM in VFA consistently across the three datasets, meaning that our model leads to fewer false alarms. This may be attributed to the built-in non-vocal detector. 

The bottom of Table \ref{tab:mdb main} shows the result of general melody extraction. The proposed method outperforms DSM \cite{BittnerDeepSalience17} and compares favorably with CRNN \cite{basaran2018CRNN}. 
In general, this suggests that our simple model is effective for both vocal melody and general melody extraction.

A closer examination of the results reveals that, compared to existing methods, our model is relatively weaker in the two pitch-related metrics, RPA and RCA, especially for MedleyDB.
For example, our model suffers from high frequency noises and make the wrong prediction sporadically. 
Detailed error analysis of our model can be found in our GitHub repo.


Table \ref{tab:mdb main} also shows that our model outperforms its ablated version almost consistently across the five metrics and the four datasets, validating the effectiveness of the non-melody detector.
Although not shown in the table, we have implemented another ablated version of our model that replaces CFP with the constant-Q transform (CQT). This would decrease the OA by about 10\% for vocal melody extraction.






\begin{table}
\centering
\begin{tabular}{|l|rrrrr|} \hline
\multicolumn{6}{|l|}{
\textbf{ADC2004 (vocal melody)}}\\
Method & VR$\uparrow$ & VFA$\downarrow$ & RPA$\uparrow$ & RCA$\uparrow$ & OA$\uparrow$\\
 \hline
DSM \cite{BittnerDeepSalience17} & \textbf{92.9} & 50.5 & 77.1 & 78.8 & 70.8 \\
Lu \& Su's \cite{lu18ismir} & 73.8 & \textbf{3.0} & 71.7 & 74.8 & 74.9\\
ours & 91.1 & 19.2 & \textbf{84.7} & \textbf{86.2} & \textbf{83.7} \\
ours (ablated)&74.3 & 6.1 & 72.0 & 75.6 & 75.1\\
\hline
%
\hline
\multicolumn{6}{|l|}{\textbf{MIREX05 (vocal melody)}}\\
DSM \cite{BittnerDeepSalience17} &\textbf{ 93.6} & 42.8 & 76.3& 77.3 & 69.6 \\
Lu \& Su's \cite{lu18ismir} & 87.3 & \textbf{7.9} & \textbf{82.2} & \textbf{82.9} & \textbf{85.8} \\
ours&84.9&	13.3&	75.4&76.6&	79.5\\
ours (ablated)&71.9&	12.6&	66.3&	67.8&	73.8\\
\hline
\hline
\multicolumn{6}{|l|}{\textbf{MedleyDB (vocal melody)}}\\
DSM \cite{BittnerDeepSalience17} & \textbf{88.4} & 48.7 &\textbf{ 72.0} & \textbf{74.8} & 66.2 \\
Lu \& Su's \cite{lu18ismir}& 77.9 & 22.4 & 68.3 & 70.0 & 70.0 \\
ours&73.7&	\textbf{13.3}&	65.5&	68.9&	\textbf{79.7}\\
ours (ablated)&62.1&	14.1&	53.1& 58.8&	68.4\\
\hline
%
\hline
\multicolumn{6}{|l|}{\textbf{MedleyDB (general melody)}} \\
DSM \cite{BittnerDeepSalience17}  &	60.9 & \textbf{24.3}& \textbf{75.1} & 69.2 & 61.7\\
CRNN \cite{basaran2018CRNN} & 69.8 &	31.0 & 71.4 &	\textbf{76.5} &\textbf{ 64.3}\\
ours&\textbf{70.9}&	26.2&	57.2&	62.5&	\textbf{64.3}\\
ours (ablated)& 66.5&	27.1&	53.3&	58.6&	59.8\\
\hline
\end{tabular}
\caption{Experiment results on several datasets. The ablated version of our model does not use the non-melody detector. The arrow next to each of the five performance metrics indicates whether the result is the higher or the lower the better. Please visit our github repo for the standard deviation values.}
\label{tab:mdb main}
\end{table}

\section{Conclusion}
\label{sec:conclusion}
We have introduced a streamlined encoder/decoder architecture that is designed for melody extraction. 
It employs only 7 convolution or up-convolution layers.
Due to the use of a built-in non-melody detector, we do not need further post-processing of the result. 
The code is public  and we hope it contributes to other music tasks.


\begin{thebibliography}{10}

\bibitem{salamon14spm}
J.~Salamon, E.~G\'{o}mez, D.~P.~W. Ellis, and G.~Richard,
\newblock ``Melody extraction from polyphonic music signals: Approaches,
  applications, and challenges,''
\newblock {\em IEEE Signal Processing Magazine}, vol. 31, no. 2, pp. 118--134,
  2014.

\bibitem{kroher16taslp}
N.~Kroher and E.~G\'{o}mez,
\newblock ``Automatic transcription of flamenco singing from polyphonic music
  recordings,''
\newblock {\em IEEE/ACM Trans. Audio, Speech, and Language Processing}, vol.
  24, no. 5, pp. 901--913, 2016.

\bibitem{bittner2017pitch}
R.~M. Bittner et~al.,
\newblock ``Pitch contours as a mid-level representation for music
  informatics,''
\newblock in {\em AES Int. Conf. Semantic Audio}, 2017.

\bibitem{beveridge18pm}
S.~Beveridge and D.~Knox,
\newblock ``Popular music and the role of vocal melody in perceived emotion,''
\newblock {\em Psychology of Music}, vol. 46, no. 3, pp. 411--423, 2018.

\bibitem{rigaud2016singing}
F.~Rigaud and M.~Radenen,
\newblock ``Singing voice melody transcription using deep neural networks.,''
\newblock in {\em ISMIR}, 2016, pp. 737--743.

\bibitem{kum2016melody}
S.~Kum, C.~Oh, and J.~Nam,
\newblock ``Melody extraction on vocal segments using multi-column deep neural
  networks.,''
\newblock in {\em Proc. ISMIR}, 2016, pp. 819--825.

\bibitem{bittner17ismir}
R.~M. Bittner et~al.,
\newblock ``Deep salience representations for $f0$ estimation in polyphonic
  music,''
\newblock in {\em Proc. ISMIR}, 2017, pp. 63--70.

\bibitem{su2018vocal}
L.~Su,
\newblock ``Vocal melody extraction using patch-based {CNN},''
\newblock in {\em Proc. ICASSP}, 2018.

\bibitem{lu18ismir}
W.-T. Lu and L.~Su,
\newblock ``Vocal melody extraction with semantic segmentation and
  audio-symbolic domain transfer learning,''
\newblock in {\em Proc. ISMIR}, 2018, pp. 521--528,
\newblock [Online] \url{https://github.com/s603122001/Vocal-Melody-Extraction}.

\bibitem{segnet}
V.~Badrinarayanan, A.~Kendall, and R.~Cipolla,
\newblock ``{SegNet}: A deep convolutional encoder-decoder architecture for
  image segmentation,''
\newblock {\em IEEE Trans. Pattern Analysis and Machine Intelligence}, vol. 39,
  no. 12, pp. 2481--2495, 2017.

\bibitem{basaran2018CRNN}
D.~Basaran, S.~Essid, and G.~Peeters,
\newblock ``Main melody extraction with source-filter {NMF} and {CRNN},''
\newblock in {\em Proc. ISMIR}, 2018.

\bibitem{BittnerDeepSalience17}
R.~M. Bittner et~al.,
\newblock ``Deep salience representations for $f_0$ estimation in polyphonic
  music,''
\newblock in {\em Proc. ISMIR}, 2017,
\newblock [Online] \url{https://github.com/rabitt/ismir2017-deepsalience}.

\bibitem{Chen2018DeepLabV3+}
L.-C. Chen, Y.~Zhu, P.~George, S.~Florian, and A.~Hartwig,
\newblock ``Encoder-decoder with atrous separable convolution for semantic
  image segmentation,''
\newblock {\em eprint arXiv:1802.02611}, 2018.

\bibitem{dong18ismir}
H.-W. Dong and Y.-H. Yang,
\newblock ``Convolutional generative adversarial networks with binary neurons
  for polyphonic music generation,''
\newblock in {\em Proc. ISMIR}, 2018.

\bibitem{southall18ismir}
C.~Southall, R.~Stables, and J.~Hockman,
\newblock ``Improving peak-picking using multiple time-step loss,''
\newblock in {\em Proc. ISMIR}, 2018, pp. 313--320.

\bibitem{su2015combining}
L.~Su and Y.-H. Yang,
\newblock ``Combining spectral and temporal representations for multipitch
  estimation of polyphonic music,''
\newblock {\em IEEE Trans. Audio, Speech, and Language Processing}, vol. 23,
  no. 10, pp. 1600--1612, 2015.

\bibitem{kobayashi1984spectral}
T.~Kobayashi and S.~Imai,
\newblock ``Spectral analysis using generalized cepstrum,''
\newblock {\em IEEE Trans. Acoust., Speech, Signal Proc.}, vol. 32, no. 5, pp.
  1087--1089, 1984.

\bibitem{tokuda1994mel}
K.~Tokuda, T.~Kobayashi, T.~Masuko, and S.~Imai,
\newblock ``Mel-generalized cepstral analysis: a unified approach to speech
  spectral estimation.,''
\newblock in {\em Proc. Int. Conf. Spoken Language Processing}, 1994.

\bibitem{su2017HSP_DNN}
L.~Su,
\newblock ``Between homomorphic signal processing and deep neural networks:
  Constructing deep algorithms for polyphonic music transcription,''
\newblock in {\em Proc. APSIPA ASC}, 2017.

\bibitem{peeters2006music}
G.~Peeters,
\newblock ``Music pitch representation by periodicity measures based on
  combined temporal and spectral representations,''
\newblock in {\em Proc. IEEE ICASSP}, 2006.

\bibitem{Klambauer17arxiv}
G.~Klambauer et~al.,
\newblock ``Self-normalizing neural networks,''
\newblock {\em arXiv preprint arXiv:1706.02515}, 2017.

\bibitem{lehner14icassp}
B.~Lehner, G.~Widmer, and R.~Sonnleitner,
\newblock ``On the reduction of false positives in singing voice detection,''
\newblock in {\em Proc. ICASSP}, 2014, pp. 7480--7484.

\bibitem{schlueter15ismir}
J.~Schl{\"u}ter and T.~Grill,
\newblock ``Exploring data augmentation for improved singing voice detection
  with neural networks,''
\newblock in {\em Proc. ISMIR}, 2015.

\bibitem{Danieljointdetect}
D.~Stoller, S.~Ewert, and S.~Dixon,
\newblock ``Jointly detecting and separating singing voice: A multi-task
  approach,''
\newblock in {\em Proc. Latent Variable Analysis and Signal Separation}, 2018,
  pp. 329--339.

\bibitem{Lin2017focal_loss}
T.-Y. Lin et~al.,
\newblock ``Focal loss for dense object detection,''
\newblock {\em eprint arXiv:1708.02002}, 2017.

\bibitem{bittner14ismir}
R.~Bittner et~al.,
\newblock ``{MedleyDB}: A multitrack dataset for annotation-intensive {MIR}
  research,''
\newblock in {\em Proc. ISMIR}, 2014,
\newblock [Online] \url{http://medleydb.weebly.com/}.

\bibitem{mireval}
C.~Raffel et~al.,
\newblock ``mir\_eval: a transparent implementation of common mir metrics,''
\newblock in {\em Proc. ISMIR}, 2014,
\newblock [Online] \url{https://github.com/craffel/mir\_eval}.

\end{thebibliography}

\end{document}